\documentclass[aps, prl, twocolumn, showpacs, superscriptaddress,longbibliography]{revtex4}
\usepackage{amssymb}
\usepackage{amsmath}
\usepackage{amsfonts}
\usepackage{amsthm}
\usepackage{bbm}
\usepackage{graphicx}

\usepackage{siunitx}
\usepackage{changes}

\def\be{\begin{equation}}
\def\ee{\end{equation}}

\begin{document}

\title{Control of the Oscillatory Interlayer Exchange Interaction with Terahertz Radiation}

\author{Uta Meyer}
\author{G\'eraldine Haack}
\thanks{Present Address: Groupe de Physique Appliqu\'ee, Universit\'e de Gen\`eve, CH-1211 Gen\`eve, Switzerland}
\author{Christoph Groth}
\author{Xavier Waintal}
\thanks{xavier.waintal@cea.fr}
\affiliation{Univ.\ Grenoble Alpes, INAC-SPSMS, F-38000 Grenoble, France}
\affiliation{CEA, INAC-PHELIQS, F-38000 Grenoble, France}
\date{\today}

\begin{abstract}
  The oscillatory interlayer exchange interaction between two magnetic layers separated by a metallic spacer is one of the few coherent quantum phenomena that persist at room temperature.
  Here we show that this interaction can be controlled dynamically by illuminating the sample (e.g.\ a spin valve) with radiation in the \SI{\operatorname{10--100}}{\tera\hertz} range.
  We predict that the exchange interaction can be changed from ferromagnetic to anti-ferromagnetic (and vice versa) by tuning the amplitude and/or the frequency of the radiation.
  Our chief theoretical result is an expression that relates the dynamical exchange interaction to the static one that has already been extensively measured.
\end{abstract}

\maketitle
% Remark J(pi/2)>0 corresponds to ferromagnetic coupling. (checked with LLG equation and relation to the energy).

The physics of magnetic multilayers at room temperature -- a system that has raised high expectations for making e.g.\ non-volatile magnetic memories or high frequency oscillators -- is essentially a semi-classical affair.
Of course, quantum mechanics is crucial to understand what happens at the atomic scale, but at larger scales phenomena such as Giant Magneto Resistance (GMR) can be understood with the Boltzmann equation (for currents in plane geometries) or even with the Drift-Diffusion equation (for currents perpendicular to plane geometries) \cite{valet1993}. Likewise, spin
torque physics \cite{slonczewski1996,berger1996} can be understood within the same theoretical framework \cite{rychkov2009}. The microscopic theory only shows up in the values of a few effective parameters (spin-polarized mean free paths, spin diffusion length, etc.) that can be either calculated or simply measured \cite{petitjean2012}.
The rationale for this theoretical status is twofold.
 First (i), it is natural to surmise that the phase coherence length $L_\phi$ (that characterizes the distance over which the phase of the electronic wave function is well-defined) is rather short: the electrons close to the Fermi surface interact with the large numbers of incoherent phonons and magnons that are present at room temperature.
 Second (ii), even if $L_\phi$ was large, a typical spin valve nanopillar involves a large number $N_{ch}$ of conducting channels (typically several tens of thousands) so that any interference effect would be typically washed out upon averaging over so many different paths \cite{waintal2000}.

These two hypotheses are a bit naive, however,  and strong quantum effects do survive at room temperature.
The typical value of $L_\phi$ can be inferred from recent experiments in Fe-MgO-Fe-MgO-Fe double tunnel junctions \cite{tao2015}. The tunneling barriers select electrons propagating mostly perpendicularly to the interfaces so that hypothesis (ii) does not hold for tunneling systems,
and one observes resonances corresponding to the Fabry-Perot interferometer formed by the two MgO layers.
The resulting $L_\phi$ is of the order of a few \si{nm} (\SI{\sim 5}{nm}).
This is not very large,
yet it is larger than the typical thickness of the normal spacers used in spin valves
(typically \SI{\operatorname{2--3}}{nm}) so that most nanopillars used in
spin torque experiments are, in fact, quantum coherent.
Going back to metallic systems, evading hypothesis (ii) is more subtle and requires distinguishing charge and spin currents. The effect of interference on the charge current is at the origin of the universal conductance fluctuations but is very small, a $1/N_{ch}$ correction. In contrast, interference effects on the spin current can remain very strong.
This fact has been known for a long time with the emblematic example of the oscillatory interlayer magnetic exchange (or RKKY) interaction.
The RKKY interaction, which is commonly used to stabilize artificial antiferromagnets, is indeed an interference effect \cite{bruno1995} as shown experimentally by its oscillatory behavior with respect to the spacer thickness \cite{parkin1990,parkin1991}.
This important observation -- the interlayer exchange interaction is a room temperature interference effect -- is key for the effects discussed below.

\begin{figure}
\includegraphics[width=\linewidth]{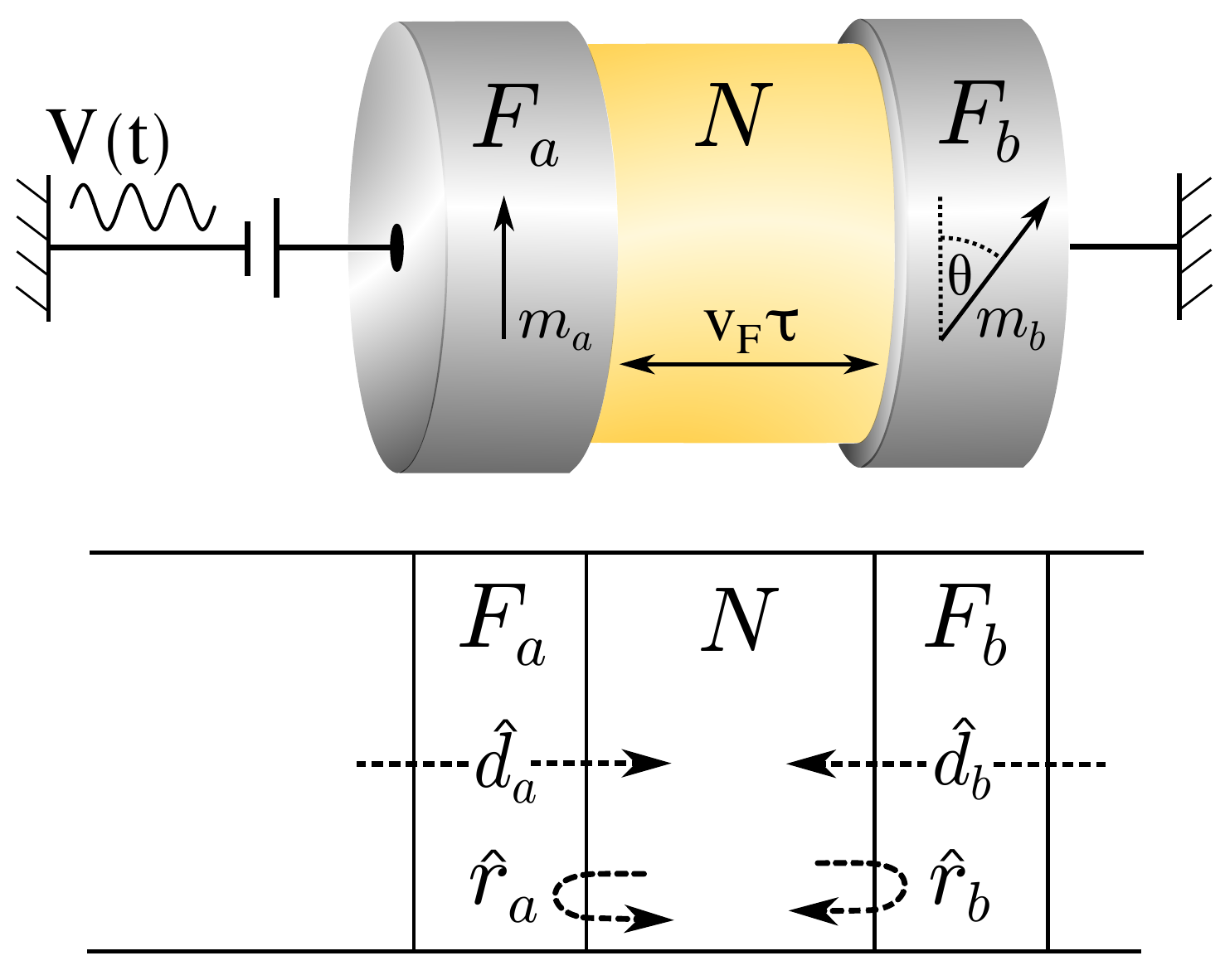}
\caption{
\label{fig:sketch}
Upper panel: schematic of a spin valve irradiated with terahertz voltage $V(t)$. The magnetizations of the two ferromagnetic layers $F_a$ and $F_b$ lie along the two unit vectors $\vec m_a$ and $\vec m_b$ that form an angle $\theta$. The characteristic time scale of the system is the time of flight $\tau$ through the normal spacer $N$ whose thickness is  $v_F \tau$ ($v_F$: Fermi velocity). Lower panel: the scattering model for the magnetic layers is defined in terms of the  transmission ($\hat{d}_a$, $\hat{d}_b$) and reflection ($\hat{r}_a$, $\hat{r}_b$) complex matrices.}
\end{figure}

In this Letter, we consider an entirely new setup consisting of a spin valve probed with terahertz radiation.
Whereas frequencies in the terahertz range were considered rather exotic only a few years ago, they are getting more and more experimental attention, including in the context of spintronics \cite{huisam2015,seifert2016}. 
This is the highest frequency band at which matter can be probed unaffected by (optical) excitations at the atomic scale.
Since the frequencies of terahertz radiation are high enough to affect directly the transport dynamics, it is beyond the common adiabatic assumption that s-like (transport) electrons see a static or adiabatically changing magnetic texture \cite{tserkovnyak2005}.
Our main finding is that this ultra-fast radiation can be used to control the interference pattern at the origin of the RKKY interaction.
We show that one can control the magnetic configuration (e.g.\ parallel versus anti-parallel) of a spin valve by varying the intensity or the frequency of incident terahertz radiation.
The underlying mechanism for this control is a dynamical modification of the energy landscape, as opposed to the damping/antidamping form of the conventional current induced spin torque.
It could potentially have distinct advantages for applications in terms of energy consumption.

{\it Scattering model for the magnetic layers.} Let us start by setting up our model for the description of the interlayer RKKY interaction.
We consider a magnetic trilayer $F_a|N|F_b$ made of a normal spacer $N$ (made of e.g.\ copper or ruthenium) sandwiched in-between two magnetic layers (e.g.\ cobalt), see Fig.~\ref{fig:sketch}. A first approach is to calculate the electronic energies of the spin valve as a function of the angle $\theta$ between the magnetization of the two magnets.
Filling the states up to the Fermi level, one obtains the energy $E(\theta)$ whose $\theta$-dependence characterizes the magnetic interaction.
Since the seminal article of Bruno \cite{bruno1995}, it is understood that the RKKY interaction can also be viewed as a scattering problem: the equilibrium spin current $\vec J$ flowing through the spacer $N$ is simply related to the RKKY interaction: $\vec J$ is perpendicular to the magnetizations of $F_a$ and $F_b$ and its value reads \cite{waintal2002},
\be
J = \frac{\partial E}{\partial \theta}.
\ee
In close analogy with the charge current, the value of $J$ can be obtained from the scattering properties of $F_a$ and $F_b$. Without loss of generality, we describe our system with one scattering matrix for each of the two magnets that accounts for both the bulk and the interface properties.
If the spacer is not fully ballistic, its scattering properties are incorporated into one of the magnetic layers which itself decomposes into reflection ($\hat r_a,\hat r_b$) and transmission ($\hat d_a,\hat d_b$) $2N_{ch}\times 2N_{ch}$ matrices.
Most of the formulas below, in particular our main result Eq.~\eqref{floquet-Lambda3} are general.
However, for some practical calculations, we ignore the channel degree of freedom and further parametrize the reflection matrices in terms of the amplitude for Majority ($r_{aM}$) and minority ($r_{am}e^{i\phi_a}$) spins where the phase $\phi_a$ accounts for the detuning between Majority and minority channels.
Introducing $\bar r_a= (r_{aM}+r_{am}e^{i\phi_a})/2$ and $\delta r_a= (r_{aM}-r_{am}e^{i\phi_a})/2$,
the reflection matrices take the simple form,
\be
\hat r_a =\bar r_a \mathbb{I}+ \delta r_a \vec\sigma \cdot \vec m_a,
\label{reflex}
\ee
where $\vec\sigma$ is the vector of Pauli matrices and $\vec m_a$ is a unit vector that lies in the direction of the magnetization of $F_a$. Similar expressions stand for the transmissions amplitudes $d_{aM}$ and $d_{am}$ as well as for the matrices that characterize the $F_b$ layer.
Neglecting spin-orbit coupling, particle/spin conservation imposes $r_{aM}^2+d_{aM}^2=r_{am}^2+d_{am}^2=1$. In practice,
the RKKY interaction is dominated by a few propagating channels (i.e.\ transverse momenta for large systems) that correspond
to specific points of the Fermi surface \cite{bruno1995,stiles1996}, so that the above parametrization has been proven to be reliable.

{\it DC RKKY theory.} The first step of the theory is to calculate the total scattering amplitudes for the two magnets in series.
The amplitude matrix $\hat \Lambda_{RL}$ of a right-going (R) mode {\it -- inside the spacer --} coming from the left (L) electrode is given by the corresponding sum over the different trajectories with 0, 1, 2, \dots{} multiple reflections \cite{waintal2000},
\be
\hat \Lambda_{RL} = \sum_{n=0}^{\infty} [\hat r_a \hat r_b e^{i 2E \tau/\hbar}]^n \hat d_a,
\label{lambda-series}
\ee
where $e^{i 2E \tau/\hbar}$ is the extra phase picked up in the
spacer ($E$: energy, $\tau$: time of flight through the normal spacer $N$). This geometrical series can be readily summed into
\be
\hat \Lambda_{RL}(E) = \frac{1}{\hat 1- \hat r_a \hat r_b e^{i 2E \tau/\hbar}} \hat d_a.
\ee
Similarly $\hat\Lambda_{LL} = \hat r_b \hat\Lambda_{RL}$.

The second step is to calculate the spin current and fill up the states, in close analogy to the Landauer formula for
the conductance \cite{waintal2000,waintal2002},
\be
\label{four-terms}
\vec J = \vec \lambda_{RL} - \vec \lambda_{LL} + \vec \lambda_{RR} - \vec \lambda_{LR}
\ee
with
\be
\label{spinlandauer}
\vec \lambda_{\alpha\beta} = \frac{1}{4\pi}{\rm Re  \ } \int dE  {\rm \ Tr  }[\hat\Lambda_{\alpha\beta}\hat\Lambda_{\alpha\beta}^\dagger \vec\sigma ]f_\beta(E)
\ee
were $f_\beta(E)$ is the Fermi function in lead $\beta$ and $\alpha,\beta\in\{ L,R\}$. We further separate the two contributions arising from the left and right electrodes:
\be
\vec J = \vec \lambda_{L} + \vec \lambda_{R},
\ee
where $\vec \lambda_{L} = \vec \lambda_{RL} - \vec \lambda_{LL}$ and $\vec \lambda_{R} = \vec \lambda_{RR} - \vec \lambda_{LR}$. The expression for $\vec \lambda_{R}$ is obtained by exchanging the role of $a$ and $b$ in the above expression for $\vec \lambda_{L}$ and multiplying by a global minus sign.
It is remarkable that the above theory captures in a single framework both the RKKY interaction as well as current-induced spin torque.
An interesting case where the calculation can be carried out explicitly is the limit of small reflections at equilibrium and zero temperature.
This limit corresponds to the textbook perturbative calculation of Friedel oscillations.
To  second order in the reflection amplitudes, we arrive (assuming for definiteness $\phi_a=\phi_b=0$) at $\vec J\propto (\vec m_a \times \vec m_b)$ with
\be\label{DC-RKKY}
J = \frac{J_0}{\pi} \delta r_a \delta r_b \cos \left(\frac{2E_F\tau}{\hbar}\right) \sin\theta,
\ee
where $J_0 = N_{ch} \hbar/\tau$ is our unit for the interlayer exchange energy and $E_F$ the Fermi energy.
Even though Eq.~\eqref{spinlandauer} involves an integral over all energies, the end result only depends on Fermi-level quantities in the (large $\tau$) limit where the energy dependence of the transmission/reflection amplitude is slow compared to $\hbar/\tau$
\footnote{This can be seen formally by expanding Eq.~\eqref{spinlandauer} in terms of the different trajectories of Eq.~\eqref{lambda-series}: each integral is essentially the Fourier transform of a product of transmission/reflection amplitudes times the Fermi function.
  At large $\tau$, this is entirely dominated by the sudden jump at the Fermi level.}.
Eq.~\eqref{DC-RKKY} is the standard expression for the RKKY interaction and explicitely shows its oscillatory character -- a hallmark of the underlying Fabry-Perot interferometer \cite{bruno1995}.
Note that even though the above simple expression
involves a number of approximations (small reflection, no disorder, no description of the Fermi surface, etc.), Eq.~\eqref{DC-RKKY} is nevertheless fairly robust \cite{stiles1996}.

{\it Time-dependent theory}. We now extend the previous theory to the time domain in order to study how the RKKY interaction is modified by a time-dependent voltage $V(t)$ applied to the left contact.
The prescription to account for the time-dependence is very simple: following \cite{gaury2014b,gaury2014a},  a phase kink
$\Phi (t) = \int_0^t dt\  eV(t)/\hbar$
propagates through the sample so that the matrix
$\hat \Lambda_{LR}$ gets modified into
\be
\label{t-Lambda}
\hat \Omega_{RL}(t) = \sum_{n=0}^{\infty} [\hat r_a \hat r_b e^{i 2E \tau/\hbar}]^n e^{-i\Phi(t-2n\tau)} \hat d_a.
\ee
The rest of the theory remains unchanged, in particular the terms $\vec \lambda_R$ coming from the right electrode are not affected by the pulses.
We immediately see from Eq.~\eqref{t-Lambda} that the interference pattern will be modified dynamically by the presence of the time dependent phase.
At this stage, we can already analyze the simple situation of the ``Dirac comb'' where a train of very narrow pulses is sent with a repetition period that exactly matches the
traveling time $2\tau$.
In that situation, the dynamical phase reduces to $\Phi(t-2n\tau) = \bar \Phi n + B$ where $\bar\Phi=\int dt eV(t)/\hbar$ is the integral over one period (a single narrow pulse) and $B$ is an irrelevant shift.
In this simple case, we recover our geometrical series
and $\hat \Omega_{RL}(t)$ is effectively time-independent:
\be
\hat \Omega_{RL} = \frac{1}{\hat 1- \hat r_a \hat r_b e^{i 2E \tau/\hbar} e^{-i\bar\Phi}} \hat d_a.
\ee
In the small reflection limit, we arrive at
\be\label{dirac-RKKY}
J = \frac{J_0}{2\pi} \delta r_a \delta r_b \left[ \cos \left(\frac{2E_F\tau}{\hbar}+ \bar \Phi\right)
+ \cos \left(\frac{2E_F\tau}{\hbar}\right)\right] \sin\theta.
\ee
Eq.~\eqref{dirac-RKKY} contains the main qualitative message of this Letter: changing the intensity of the pulses tunes $\bar\Phi$ which in turns controls the intensity of the RKKY interaction.
One can easily go from a situation where $J$
favors a parallel configuration to an antiparallel one, switch off the interaction (case where $\bar \Phi =\pi$), or on the contrary switch it on if it was absent in the initial configuration (case where $4E_F\tau/h$ is an integer).

{\it Floquet theory.} We now proceed with the general case and suppose that $V(t)$ is periodic with frequency $\omega$.
We can expand the dynamical phase in the Fourier series
\be
e^{-i\Phi(t)} = e^{-i\frac{\bar\Phi\omega t}{2\pi}}\sum_{p} P_p e^{-i\omega p t},
\ee
where $\bar\Phi$ and the Fourier coefficients $P_p$ fully characterize the pulses.
We arrive at
\be
\label{floquet-Lambda2}
\hat \Omega_{RL}(t) = \sum_{p}  \hat \Lambda_{RL}\left(E -\frac{\hbar\omega\bar\Phi}{2\pi} -p\hbar\omega \right) e^{i\omega pt},
\ee
from which, together with Eqs.~\eqref{four-terms} and \eqref{spinlandauer},  we  calculate the effect of an arbitrary pulse shape.
Focusing on the part of the RKKY interaction that does not oscillate in time (for the applications considered here the oscillating part is too fast to have a significant impact on the magnetization -- the situation might be different in antiferromagnets which can possess \si{THz} resonances \cite{Jin2013}), we get
\be
\label{floquet-Lambda3}
\vec J = \vec\lambda_R(E_F) +\sum_p P_p^2 \ \vec\lambda_{L}\left(E_F-\frac{\hbar\omega\bar\Phi}{2\pi} -p\hbar\omega \right).
\ee
Eq.~\eqref{floquet-Lambda3} is the chief result of this Letter.
It relates the RKKY interaction in presence of an arbitrary pulse to  the
DC RKKY interaction.
From now on, we focus on the case of sinusoidal radiation $V(t) = V_0 \sin\omega t$ because of its experimental relevance \cite{Wang2015}.
The coefficients $P_p$ are given by the Bessel function of the first kind: $P_p = J_p(eV_0/\hbar\omega)$.
Typically $P_p$ is maximal for $p\approx eV_0/\hbar\omega -1$ and decays at large $p$ while $\sum_p P_p^2 = 1$.
For instance, for $eV_0/\hbar\omega = 1.8$, we have $P_0^2=0.12$,
$P_1^2=0.68$, $P_2^2=0.19$, so that the linear combination of Eq.~\eqref{floquet-Lambda3} is dominated by the $p=1$
harmonic whose contribution can be tuned by changing the frequency $\omega$.

\begin{figure}
\includegraphics[width=\linewidth]{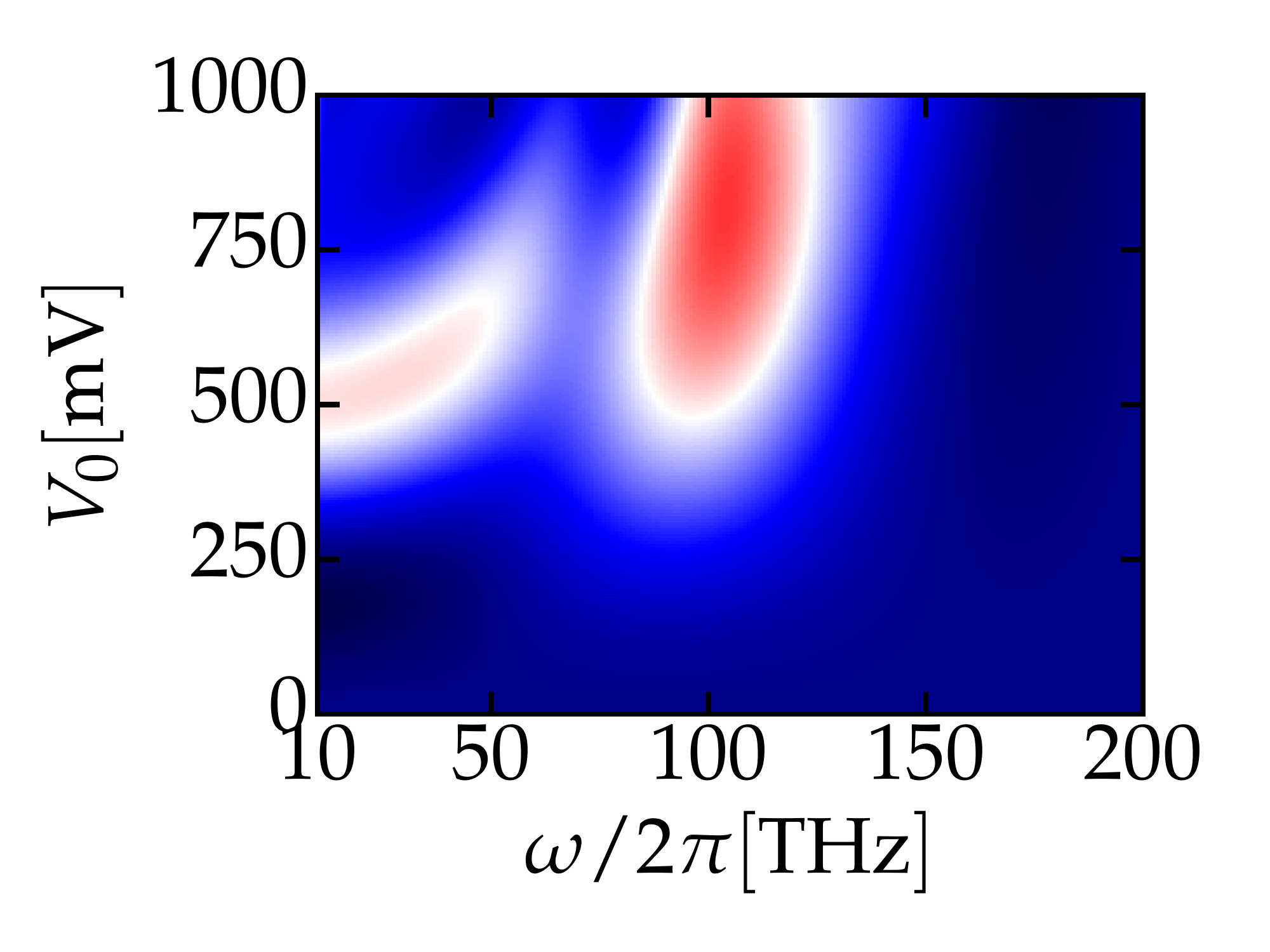}
\caption{
\label{color_zhang}
Interlayer exchange energy $J(\theta=\pi/2)$ as a function of the frequency $\omega$ and amplitude $V_0$ of the terahertz radiation.
The dc exchange interaction is described by Eq.~\eqref{eq:zhang} and corresponds to the Co-Ru-Co system of \cite{zhang1994} with a Ru thickness of $2.4$  nm.
The color code goes from \SI{-0.05}{mJ/m^2} (antiferromagnetic coupling, dark blue) to \SI{+0.05}{mJ/m^2} (ferromagnetic coupling, red).}
\end{figure}

An important aspect of Eq.~\eqref{floquet-Lambda3} is that it is very general and holds for arbitrary (disordered) interfaces and Fermi surfaces.
We can now rely on our existing knowledge of the equilibrium RKKY to predict the effect of a given pulse.
To this end, one possibility is to use ab-initio calculations such as \cite{stiles1996}. In the case of symmetric spin valves ($F_a = F_b$) where $\lambda_R=\lambda_L = J_{eq}/2$, a very appealing possibility is to
simply measure $J_{eq}$ experimentally (by varying the thickness of the metallic spacer) without resorting to further theory.
The corresponding oscillations have been observed in numerous experiments (see for instance Fig.~3 of \cite{parkin1990}).
To illustrate this strategy, let us focus on Fig.~16
of Ref.~\cite{zhang1994} where a careful analysis of the oscillations of the saturation field was used to extract the RKKY interaction.
From the period and amplitude of the oscillations, we find that (assuming a Fermi velocity $v_F = \SI{10^6}{m/s}$) the equilibrium
RKKY interaction of the Co-Ru-Co multilayer of \cite{zhang1994} is reasonably well described by
\be
\label{eq:zhang}
J_{eq} \approx \left( \frac{A_0}{v_F\tau} \cos (2 E_F \tau/\hbar) + \frac{A_1}{v_F\tau} \cos (4 E_F \tau/\hbar) \right) \sin\theta
\ee
with $A_0\approx 3 \times \SI{10^{-13}}{J.m}$,
 $A_1\approx \SI{10^{-13}}{J.m}$
 and $\tau$ between \SI{1}{fs} and \SI{2.5}{fs}. Eq.~\eqref{eq:zhang} can now be inserted into Eq.~\eqref{floquet-Lambda3} to predict the effect of terahertz radiation.
 The result is shown in Fig.~\ref{color_zhang} for an antiferromagnetic sample.
 We find that the radiation can actually be used to reverse the sign of the exchange interaction from antiferromagnetic to ferromagnetic
 (red pockets).

\begin{figure}
\includegraphics[width=\linewidth]{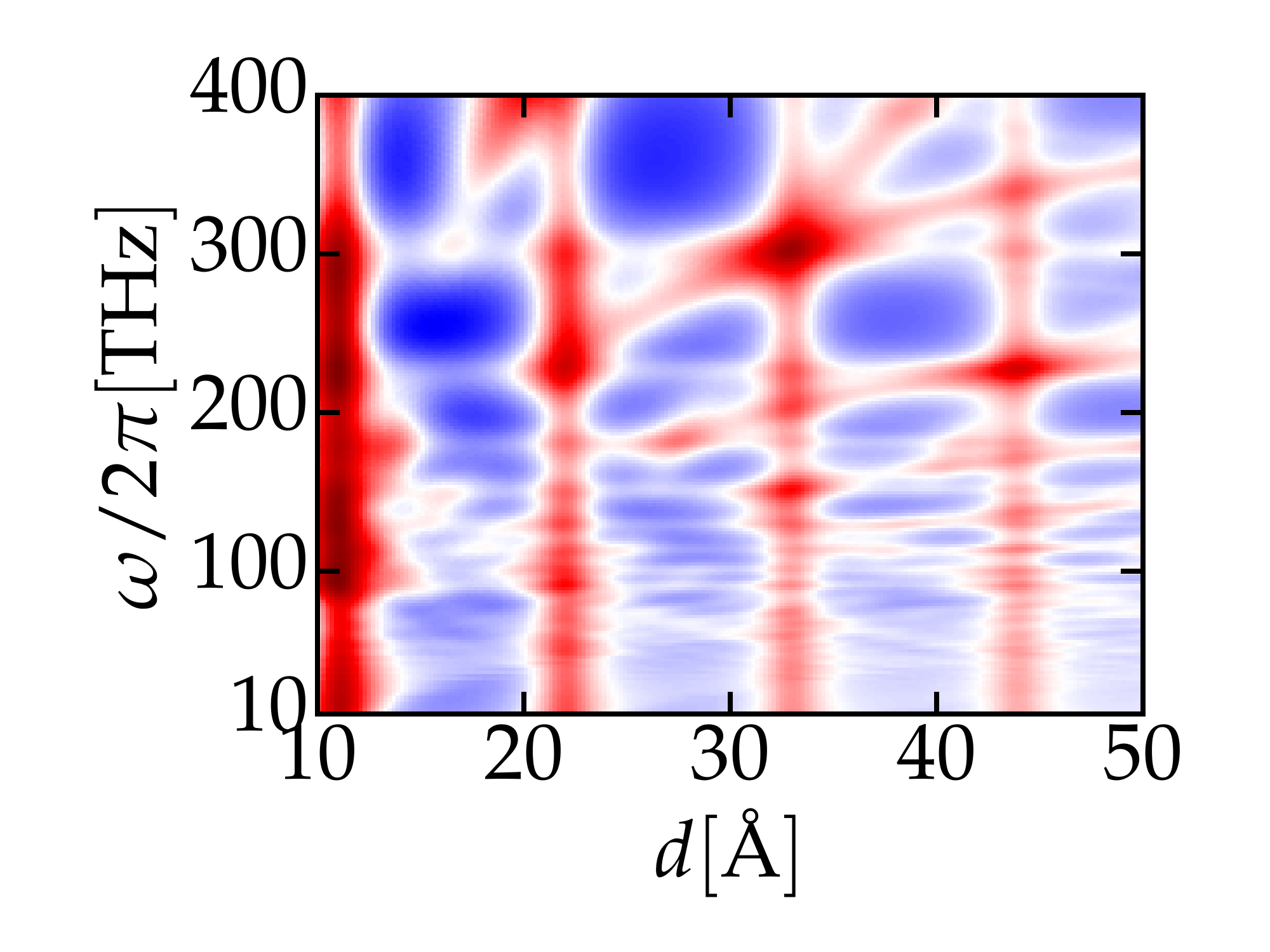}
\caption{
\label{fig3}
Example of numerical calculation of the exchange energy $J(\theta=\pi/2)$ as a function of the spacer thickness $d$ (in Angstrom) and frequency (in \si{THz}) for a voltage amplitude $V_0 = \SI{100}{mV}$.
The parameters used for the calculation were
$|d_{aM}|^2=0.41$, $|d_{am}|^2=0.22$, $|d_{bM}|^2=0.83$, $|d_{bm}|^2=0.59$, $\lambda_F = \SI{11}{\angstrom}$ and $v_F = \SI{10^6}{m/s}$.
The color code indicates antiferromagnetic coupling (red) and ferromagnetic coupling (blue) with a maximum intensity of the order of $\SI{1}{mJ/m^2}$.}
\end{figure}

 The situation gets even more interesting in the case of asymmetric structures: in the case where $|\lambda_R|\ll |\lambda_L|$, the RKKY is entirely dominated by the radiation.
 In order to
 evaluate the two contributions $\vec\lambda_R$ and $\vec\lambda_L$ numerically, we define the energy dependence of the reflection parameters $r_{aM}(E)$, $r_{am}(E)$, \dots{} so that they vary slowly with respect to $\hbar/\tau$. In this limit, the end results depend only of $r_{aM} \equiv r_{aM}(E_F)$, $r_{am} \equiv r_{am}(E_F)$, \dots{} irrespective of the actual energy dependence (we have explicitly checked this point numerically for various energy dependences). A typical result is shown in Fig.~\ref{fig3}. The structure of the thickness-frequency (or frequency-voltage) map is fairly rich and can depend significantly on the microscopic parameters.
 However, the salient features are very robust: in all systems, one can reverse the sign of the RKKY coupling from antiferromagnetic to ferromagnetic and vice versa (except in totally symmetric systems at the particular thicknesses where the RKKY is at its strongest).

{\it Discussion.} The physics discussed in this Letter is controlled by one important time scale, the time of flight $\tau$ through the metallic spacer.
The typical thicknesses of interest lie between \SI{1}{nm} and \SI{10}{nm}, so that (assuming a Fermi velocity $v_F \approx \SI{10^6}{m/s}$) $\tau$ lies between \SI{1}{fs} and \SI{10}{fs}.
Since the relevant frequency range is significantly smaller than $2h/\tau$ we arrive at frequencies in the \SI{\operatorname{10--100}}{THz} range, i.e.\ in the terahertz/far infrared regime.
In terms of amplitudes, the non-perturbative effects described here are at their strongest for $eV_0/\hbar\omega \approx 1$ which translates into $V_0$ in the \SI{\operatorname{50--1000}}{meV} range.
Assuming that the terahertz source impedance can be matched with the one of the sample (not necessarily an easy task), our proposal requires a \SI{\operatorname{6--1200}}{\micro\watt} source that is well possible with current technology (tens of \si{mW} with dc quantum cascade lasers and hundreds of \si{mW} with pulsed versions, see \cite{Bachmann2016, Vitiello2015, Wang2015} and references therein).

To conclude, we have shown that interlayer exchange interaction can be dynamically modified by terahertz radiation.
This effect is potentially promising as a new means to achieve magnetic reversal, but also as a new spectroscopic tool to probe the dynamics of s-like electrons in itinerant magnetic systems \cite{Frietsch2015}.
A particularly appealing aspect of the above theory is
that almost all effects that could destroy the RKKY interaction (disorder, interface mismatch, decoherence, etc.) are equally
harmful to both the equilibrium case and the dynamical case that we have investigated in this Letter.
Therefore, the robustness of the RKKY interaction that has been observed experimentally over the years is a strong indication that the present effects should
be amenable to experimental observation.

{\it Acknowledgements} We thank J.-M.~Gerard, T.~Valet and S.~Bourdel for valuable discussions. G.~H. acknowledges support from the Swiss National science fundation through the Marie Heim-V\"ogtlin grant no.\ 18874. X.~W. and C.~G. acknowledge support from the ANR QTERA.

\bibliography{pulsed-rkky_ref}

\end{document}